\documentclass[sigconf,nonacm]{acmart}

\usepackage{soul}
\usepackage{enumitem}
\usepackage{etoolbox}
\usepackage{makecell}
\usepackage{hhline}
\usepackage{tabularx}
\usepackage{algorithm}
\usepackage{algorithmic}
\usepackage{graphicx}
\usepackage{subcaption}
\usepackage{listings}
\usepackage{amsmath}
\usepackage{balance}
\usepackage{xcolor}
\usepackage[table]{xcolor}

\newcolumntype{V}{>{\centering\arraybackslash}X}

\lstdefinelanguage{ERQL}{
    morekeywords={CREATE, ABSTRACT, ENTITY, RELATIONSHIP, IMPLEMENTS, ALTER, SET, ROOT, REFERENCES, RETENTION, INTERVAL, OVERRIDING, SELECT, FROM, WHERE, FILELIST, SNAPSHOT, DELTA},
    sensitive=true,
    morecomment=[l]{--},
    morecomment=[s]{/*}{*/},
    morestring=[b]',
    morestring=[b]"
}

\BeforeBeginEnvironment{lstlisting}{\lstset{basicstyle=\ttfamily\footnotesize}}
\AfterEndEnvironment{lstlisting}{\lstset{basicstyle=\ttfamily}}

\begin{document}

\title{Towards Unified File Management for Data Lakehouse Systems}

\author{Keonwoo Oh}
\affiliation{%
    \institution{University of Maryland}
    \city{College Park}
    \state{Maryland}
    \country{USA}}
% \email{koh3@umd.edu}

\author{Akshay Mehra}
\affiliation{%
    \institution{Dolby Labs.}
    \city{Atlanta}
    \state{Georgia}
    \country{USA}}

\author{Joshua Kimball}
\affiliation{%
    \institution{Dolby Labs.}
    \city{Atlanta}
    \state{Georgia}
    \country{USA}}

\author{Amol Deshpande}
\affiliation{%
    \institution{University of Maryland}
    \city{College Park}
    \state{Maryland}
    \country{USA}}
% \email{amol@umd.edu}

\renewcommand{\shortauthors}{Oh et al.}

\begin{abstract}

Data lakehouse systems have garnered widespread adoption, yet their metadata management systems face a fundamental architectural dilemma. File-based OTFs (Open Table Formats) scale well, but suffer from metadata bloat, high commit latencies, the absence of multi-table transactions, and unsafe garbage collection. On the other hand, catalogs relying entirely on a transactional database incur severe scalability bottlenecks due to analytic file-level metadata. To reconcile this tension, we present Raven, an operational catalog that stores transactional metadata within a database while offloading analytical metadata to underlying OTF structures. Based on this hybrid architecture, Raven introduces an extended Merge-on-Read (MoR) mechanism to reduce commit latency and mitigate metadata bloat, alongside multi-table transactions under snapshot isolation. Furthermore, Raven unifies core data lakehouse semantics including versioning and reference tracking under an extensible ER (Entity-Relationship) metadata model, enabling safe global garbage collection across diverse types of assets. Empirical evaluations demonstrate that Raven substantially reduces commit latency and metadata bloat when integrated with Apache Iceberg and Delta Lake, maintains a catalog storage footprint orders of magnitude smaller than database-only systems like DuckLake, and achieves scalable garbage collection with minimal disruption.
\footnote{Code is available at: \url{https://github.com/umddb/raven}}

\end{abstract}




% Strips "ACM Reference Format" block from the first page
\settopmatter{printacmref=false}

\maketitle

\section{Introduction}
\label{sec:introduction}

\begin{figure}[t]
  \centering
  \includegraphics[width=\linewidth]{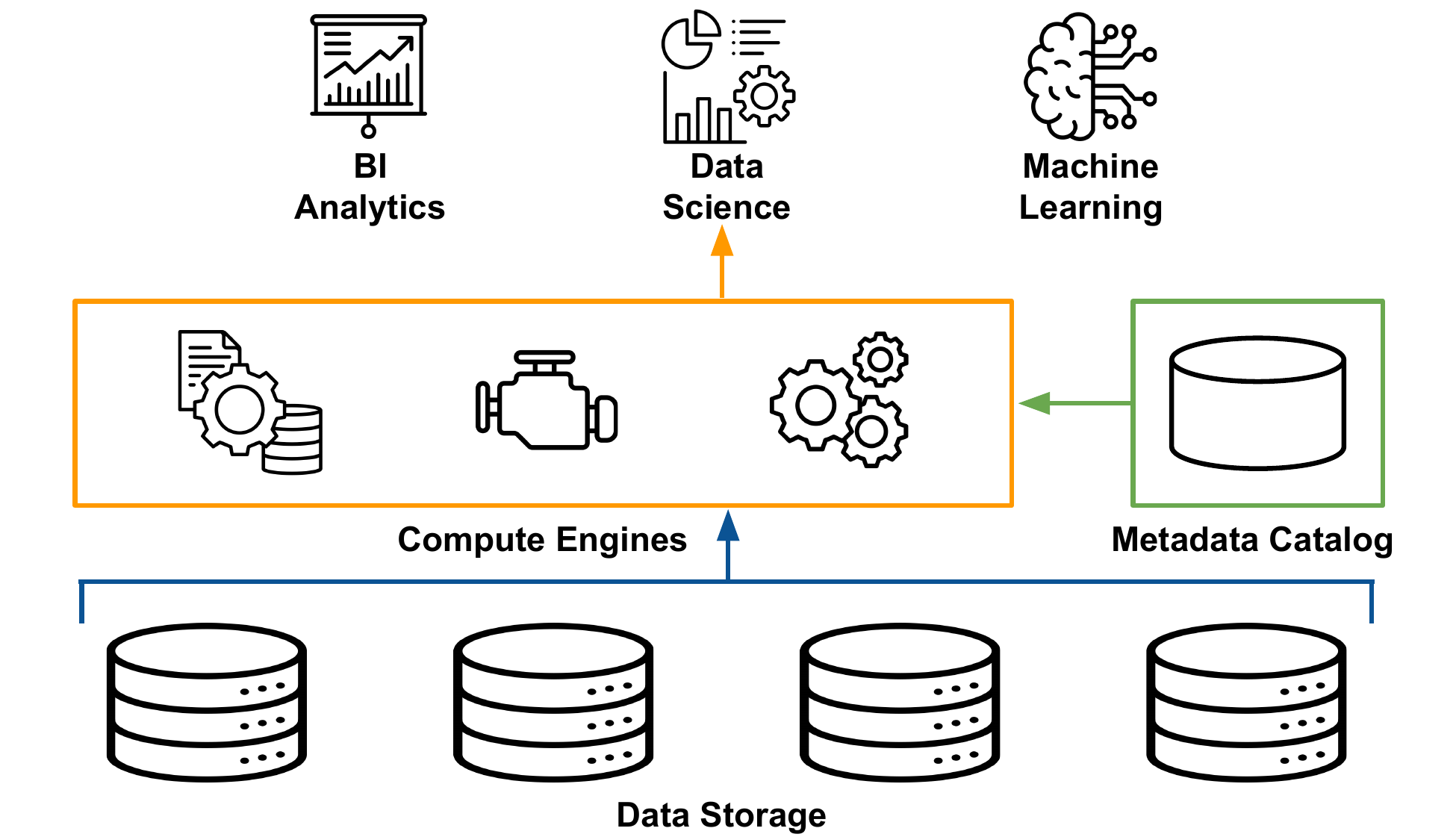}
  \caption{An example data lakehouse architecture, characterized by disaggregated components/subsystems.}
  \label{fig:datalakehouse}
\end{figure}

Historically, big data processing engines and traditional data warehouses served distinct roles~\cite{friendsorfoes}. However, the data lakehouse architecture, designed to combine the functionalities of both system types, has become more commonplace recently~\cite{armbrust2021lakehouse, datavirtualization}. Unlike traditional systems that manage data in isolation, a data lakehouse achieves a composable data stack where multiple engines operate directly on shared datasets in scalable storage (e.g., cloud object stores) using standard formats like Apache Parquet~\cite{parquet} and ORC~\cite{orc} (Figure~\ref{fig:datalakehouse}). Simultaneously, it aims to support traditional warehouse functions such as transactions, data maintenance, and access control.

Increasing adoption of the data lakehouse architecture is closely tied to the emergence of advanced metadata systems. On the one hand, there are lakehouse storage formats, also known as OTFs (Open Table Formats), including Delta Lake~\cite{deltalake}, Apache Iceberg~\cite{apacheiceberg}, and Apache Hudi~\cite{apachehudi}; these specify how to store the metadata of a tabular dataset, such as the table schema, partitioning scheme, list of data files, and file-level statistics, as a set of files alongside the data. On the other hand, a data lakehouse also requires a separate catalog service that provides a mapping from the logical namespace (e.g., database and table names) to the file directory path(s) as well as data governance and access control functions. 

Despite growing popularity, data lakehouses face significant limitations in their metadata systems. We examine the traditional data warehouses and earlier big data platforms before discussing the limitations of data lakehouse metadata systems.

\subsection{Data Warehouses and Big Data Systems}

Commercial data warehouses, including Snowflake~\cite{snowflakeelastic}, Google BigQuery~\cite{biglake}, Amazon Redshift~\cite{redshiftreinvented}, Microsoft Fabric~\cite{fabric}, ClickHouse~\cite{clickhouse}, and SingleStore~\cite{singlestore}, typically rely on transactional systems or distributed coordination services for metadata management. For example, Snowflake uses FoundationDB~\cite{foundationdb}, a highly scalable transactional key-value store, for metadata storage~\cite{motivala_2018}. However, closed architectures leave many advanced features unavailable for external data assets~\cite{snowflakeiceberg,bigqueryexternal}.

On the other hand, big data processing engines such as Hive~\cite{thusoo2010hive}, Spark~\cite{armbrust2015spark}, Presto/Trino~\cite{sethi2019presto}, Impala~\cite{impala}, and Flink~\cite{carbone2015apache} have long used the HMS (Hive Metastore) as their primary operational catalog. An HMS instance consists of a Thrift server, a DataNucleus ORM (Object-Relational Mapping) layer, and an RDBMS backend. Because the RDBMS maintains only coarse-grained mappings from tables or partitions to directory paths, query engines must retrieve these paths from HMS and issue \textit{list()} calls against the file system (cloud object store) to locate data files. For large tables with millions of files in cloud object stores, this introduces severe latency bottlenecks~\cite{armbrust2020delta}. In addition, lacking file-level metadata, HMS does not support transactional guarantees under concurrent operations.

\begin{table}[t]
  \caption{PostgreSQL storage size for DuckLake when scaling integer column count in a single table with 100,000 data files.}
  \label{tab:ducklake_metadata_size}
  \centering
  \begin{tabular}{lcccc}
    \toprule
    \textbf{Column Count} & 10 & 100 & 1,000 & 10,000 \\
    \midrule
    \textbf{Metadata Storage (GB)} & 0.15 & 1.14 & 10.97 & 109.31 \\
    \bottomrule
  \end{tabular}
  
\end{table}

\begin{table}[t]
\centering
\caption{Metadata placement across three representative data lakehouse metadata systems.}
\label{tab:metadata_placement}
\footnotesize
\setlength{\tabcolsep}{3.2pt}
\renewcommand{\arraystretch}{1.15}
\begin{tabular}{lccc}
\toprule
\textbf{Metadata Type} & \textbf{HMS + Iceberg} & \textbf{Raven + Iceberg} & \textbf{DuckLake} \\
\midrule
Table Name       & Database      & Database      & Database \\
OTF Root Path    & Database      & Database      & N/A \\
Table Versioning & Object Store & Database      & Database \\
Table Schema     & Object Store & Database      & Database \\
File Visibility  & Object Store & Database      & Database \\
\rowcolor{gray!18}
\textbf{File Statistics} & \textbf{Object Store} & \textbf{Object Store} & \textbf{Database} \\
\bottomrule
\end{tabular}
\end{table}

\subsection{Open Table Formats and Hidden Cost of Scalability}
\label{subsec:otfs}

OTFs, including Delta Lake, Apache Iceberg, and Apache Hudi, were created to address the limitations of HMS. By materializing file-level metadata as physical files alongside data, OTFs avoid storage system's expensive  \textit{list()} operations. They also support single-table ACID transactions based on OCC (Optimistic Concurrency Control) mechanisms and advanced features such as time-travel, branching/cloning, schema evolution, and maintenance operations. Meanwhile, modern catalog systems, such as Unity Catalog~\cite{unity}, Polaris~\cite{polaris}, OpenHouse~\cite{openhouse}, OneHouse~\cite{onehouse}, and Nessie~\cite{nessie} have emerged to replace HMS. However, like HMS, these catalogs store a pointer to the root directory or metadata file for each table and delegate file-level metadata tracking to the underlying OTFs.

Delta Lake and Hudi use log-based mechanisms to record file additions, deletions, and schema changes in ordered log files, reconstructing snapshots by replaying logs from checkpoints. On the other hand, Iceberg uses a hierarchical structure: a root metadata JSON file points to manifest list Avro files, which in turn reference manifest Avro files containing individual data file paths and file-level statistics.

The guiding design principle behind OTFs is that metadata is no longer small~\cite{edara2021big,hudi_lakehouse}. To manage scalability, they offload most metadata as files into cloud object store, ignoring that many metadata operations remain transactional and advanced ACID guarantees are necessary to support complex features with data integrity. We illustrate these limitations using Iceberg's \textit{appendFiles()} method. Even adding a single data file (any data modification) results in three new nested metadata files: a manifest file, a manifest list file, and a root metadata file, causing severe bloat. The commit protocol concludes with a single compare-and-swap (CAS) operation on the catalog reference to the root metadata file, inherently restricting ACID guarantees to a single table. This process incurs high commit latency due to multiple file writes and the final CAS operation. 

We formalize the main limitations of OTFs as follows:

\begin{enumerate}[leftmargin=*,noitemsep,topsep=0pt]
    \item \textbf{Small Files Problem:} Frequent transactional updates (e.g., from streaming data) can create a large number of small metadata files, degrading read performance and straining the storage system's metadata layer~\cite{iceberg_files}. 
    
    \item \textbf{High Commit Latency:} The commit protocol involves writing new metadata files, followed by a final CAS operation, which together introduce overhead due to high latencies of cloud object stores~\cite{jain2023analyzing,treecat}.

    \item \textbf{No Multi-table Transactions:} OTFs support only single-table transactions as they rely on a single CAS operation on a table-level object for the final commit~\cite{armbrust2020delta,hudi_concurrency}.
    
    \item \textbf{Incorrect File Deletion:} Updates, compaction, etc., can result in obsolete files, making garbage collection essential, but because metadata is stored as files in cloud object store, reference tracking is expensive and difficult to represent. Current systems fail to keep track of multiple access paths, so files in active use may be deleted (Section~\ref{subsec:otfgclimit}).
\end{enumerate}

\noindent Ecosystem fragmentation presents another major challenge as each OTF relies on ad-hoc workarounds and custom semantics to support similar features like transactions, branching, and garbage collection. Consequently, as feature sets expand, system complexity grows rapidly.

\subsection{Centralization and Big Metadata Problem}
\label{subsec:bigmeta}
Recent systems addressing these issues, such as DuckLake~\cite{ducklake} and TreeCat~\cite{treecat}, represent the opposite extreme by storing all metadata in a transactional database. DuckLake utilizes an RDBMS (e.g., PostgreSQL, MySQL, or DuckDB) as its backend, whereas TreeCat relies on a custom hierarchical transactional store. While these systems resolve the above OTF limitations, they encounter scalability bottlenecks already observed in industry. For example, Google BigQuery, despite using Spanner, had to offload a subset of file-level metadata into a BigQuery table~\cite{edara2021big}. Databricks recently reported compressed metadata for a single table reaching up to 1~TB~\cite{xin2026three}.

The primary culprit behind this scalability bottleneck is file-level statistics, which are necessary for cost-based query optimization and data pruning. These statistics track at least four values (row count, null count, minimum, and maximum) per column per file, increasing linearly with the product of column count and file count. For wide tables (1,000s of columns) spanning millions of files, this creates a severe scalability challenge. We empirically verify this by measuring DuckLake's metadata footprint (RDBMS size) across increasing column counts for a table with 100,000 data files (Table~\ref{tab:ducklake_metadata_size}). At 10,000 columns, the footprint reaches 109.31~GB, with statistics accounting for over 99.9\% of it. Across thousands of tables, maintaining the cumulative metadata inside a transactional store becomes difficult to scale.

\subsection{Raven: An Extensible Hybrid System}
Metadata operations inherently encompass both \textbf{transactional} and \textbf{analytical} workloads. Hence, relying exclusively on either cloud object storage or a transactional database is suboptimal. At the same time, unifying common functionality while preserving extensibility is vital for long-term system evolution. In this paper, we present Raven, a standalone catalog system with two main contributions:

\begin{enumerate}[leftmargin=*,noitemsep,topsep=0pt]
\item \textbf{Hybrid Architecture:} Raven pulls the subset of metadata requiring transactional guarantees into a database, while retaining most metadata in standard OTFs for scalability (Table~\ref{tab:metadata_placement}). We implement two features, extended Merge-on-Read (MoR) and multi-table transactions, that address the first three OTF limitations in Section~\ref{subsec:otfs} while avoiding main scalability bottlenecks of a database-only design.

\item \textbf{Unified Extensible Interface:} We unify core lakehouse semantics, such as versioning and reference tracking, through an ER metadata model that enables safe global garbage collection (addressing the last OTF limitation). Anticipating future lakehouse formats (not necessarily tabular data), Raven provides an extensible abstraction designed to generalize across diverse asset types.

\end{enumerate}

\noindent In the next section, we describe Raven's core metadata model and system interface.

\section{Metadata Model and System Interface}
\label{sec:model}

Storing critical metadata in a unified transactional store requires balancing control and extensibility. Excessive control results in a rigid system, while too little fragments the ecosystem with redundancies. To address this, we introduce three important constructs. We now cover the main ER model and the \verb|FILELIST| construct, while deferring the discussion of garbage collection semantics to Section~\ref{subsec:refrel}.

\begin{figure}[t]
  \centering
  \includegraphics[width=\linewidth]{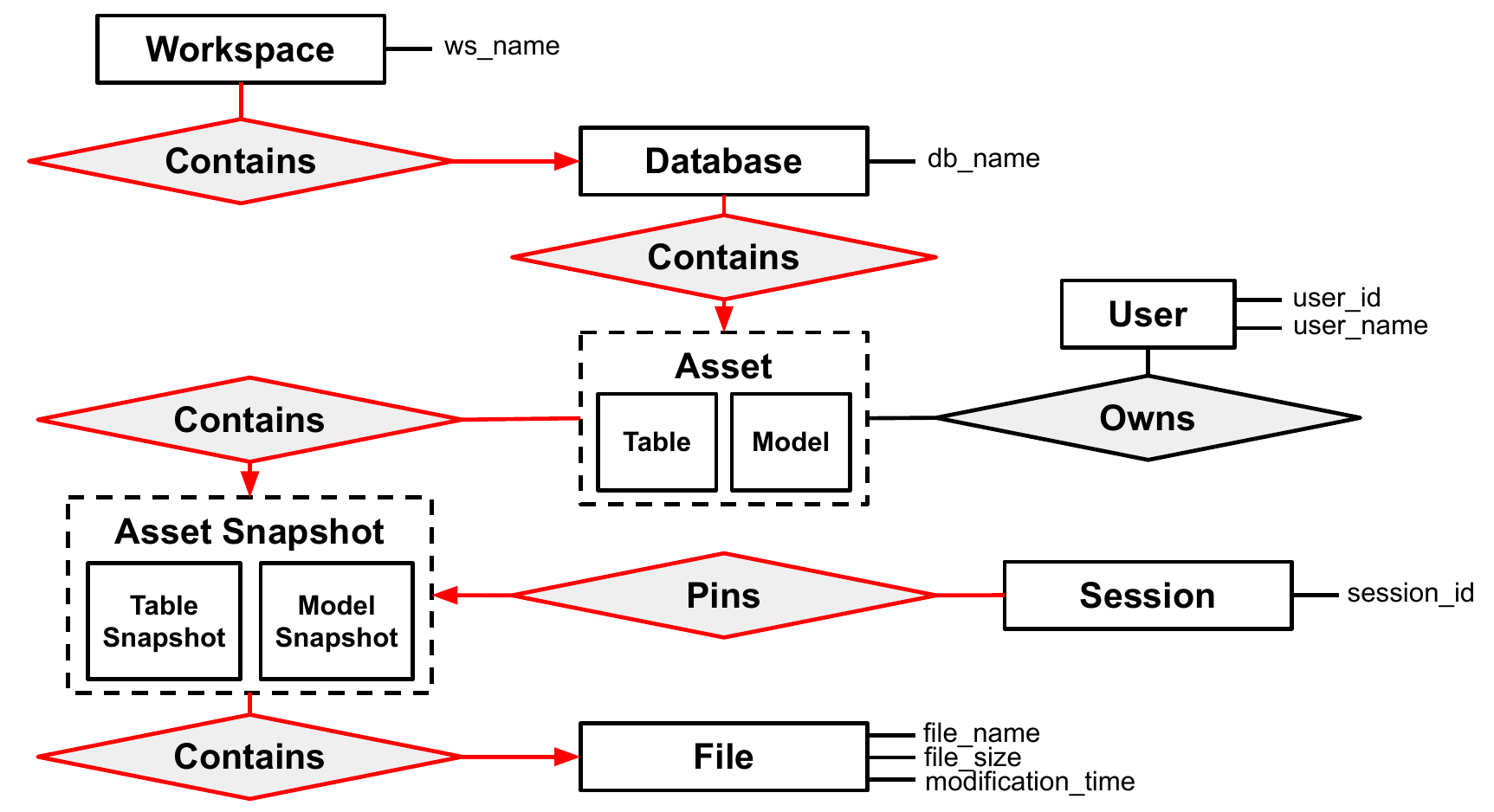}
  \caption{An example metadata schema in the ER representation. Referential relationships (Section~\ref{subsec:refrel}) are colored in red, and abstract entities are marked by dashed boundary lines.}
  \label{fig:erdiagram}
\end{figure}

\subsection{Entity-centric Abstraction}

In addition to file-level metadata, catalog metadata includes logical entities (e.g., databases, tables, and users) and their relationships. To represent them, we adopt a modified ER (Entity-Relationship) model~\cite{er}. We introduce a DSL (Domain-Specific Language) called ERQL (Entity-Relationship Query Language) with a unified syntax: a DDL (Data Definition Language) for defining and extending schemas, and a DML (Data Manipulation Language) for catalog operations. Below is DDL for the schema in Figure~\ref{fig:erdiagram}.

\smallskip

\begin{lstlisting}
CREATE ENTITY Workspace(ws_name VARCHAR(255));
CREATE ENTITY Database(db_name VARCHAR(255));
CREATE RELATIONSHIP Contains(Workspace, Database);
\end{lstlisting}

\smallskip

\noindent A highly demanded extension for the catalog is the registration of non-tabular asset types, such as machine learning models and multimedia data. To support this, we introduce the \verb|ABSTRACT ENTITY| type. It specifies the \verb|RELATIONSHIP| types available to its derived \verb|ENTITY| types while intentionally leaving the attribute schema unconstrained. As shown below, distinct asset types can be defined as derived entities of the \verb|Asset ABSTRACT ENTITY| with their own attribute schemas while inheriting shared relationships like \verb|Contains(Database, Asset)| as shown below.

\begin{lstlisting}
CREATE ABSTRACT ENTITY Asset();
CREATE RELATIONSHIP Contains(Database, Asset);
CREATE ENTITY Table (...) IMPLEMENTS Asset;
CREATE ENTITY Model (...) IMPLEMENTS Asset;
\end{lstlisting}

\smallskip

\noindent In the system execution path, the DML layer sits between the top-level catalog APIs and the underlying storage backends. We illustrate how catalog operations can be mapped to the ERQL DML. Consider \textit{getTable()}, which retrieves the table-level metadata for a specific workspace, database, and table name. The operation can be directly mapped to the following parameterized ERQL query:  

\smallskip

\begin{lstlisting}
SELECT t.*
FROM Contains(Workspace w, Database d),
     Contains(d, Table t)
WHERE w.ws_name=$1 AND d.db_name=$2 AND t.name=$3;
\end{lstlisting}

\smallskip

\noindent Each entity is uniquely identified by its type and object ID and can also be queried by the object ID.

\subsection{FileList Entity}

To address the scalability challenges of file-level metadata, we introduce the \verb|FILELIST| construct, which models any asset as a versioned collection of files. It addresses two primary challenges: (1) storing file-level statistics (Section~\ref{subsec:bigmeta}), and (2) tracking file references as standard relationships in the ER model. For example, even when a new asset snapshot differs only slightly from its predecessor, a naive ER representation must create millions of duplicate file references.

We resolve the first challenge by separating transactional and analytical file-level metadata. \verb|FILELIST| retains only the essential attributes required for transactional correctness (e.g., file path and format), while offloading file-level statistics, unnecessary for determining file visibility and lifetime, to underlying OTF structures. Meanwhile, the catalog transactionally tracks the OTF metadata files themselves within the \verb|FILELIST|, ensuring \textbf{all files} remain under its control. To resolve the second challenge, we avoid modeling every snapshot-to-file association as an explicit ER relationship, relying instead on the constrained relational representation of \verb|FILELIST|.

\begin{figure}[t]
  \centering
  \includegraphics[width=\linewidth]{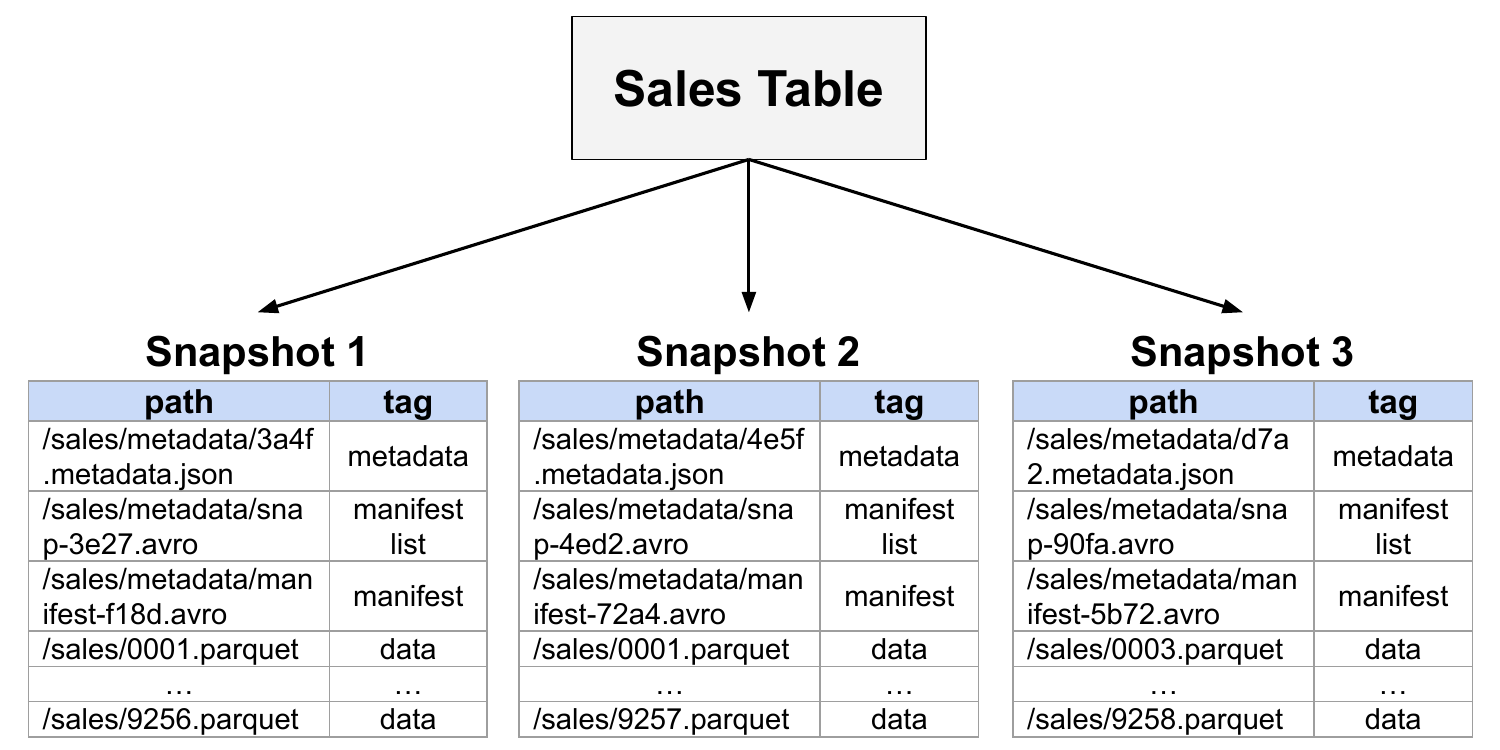}
  \caption{Consecutive Iceberg table snapshots are represented as FILELIST versions. A few attributes of the FILELIST view are omitted for brevity.}
  \label{fig:filelist}
\end{figure}

\verb|FILELIST| represents a collection of files, uniquely identified by an object ID and a version ID. A chain of \verb|FILELIST| versions sharing the same object ID forms a snapshot history, with each version referencing its file collection. For example, a table snapshot containing all files visible at a given time can be declared a \verb|FILELIST| entity type:

\begin{lstlisting}
CREATE FILELIST ENTITY TableSnapshot (table_name, create_time,...);
\end{lstlisting}

\noindent The collection of files is represented as a \textbf{relational view} with a fixed schema containing only basic file attributes such as file path, format, size, modification time, and tag. 

As an example, we show how an Iceberg table is represented as a \verb|FILELIST| in Raven. Although an Iceberg snapshot consists of a hierarchy of metadata and data files (Section~\ref{subsec:otfs}), \verb|FILELIST| provides a flat relational view of all constituent files regardless of internal semantics (Figure~\ref{fig:filelist}). Each \verb|FILELIST| version captures a specific snapshot, referencing new metadata and data files while sharing unmodified files with other versions. For example, the query below retrieves all data files for object ID 13 and version ID 2:

\begin{lstlisting}
SELECT file_path, file_size, format 
FROM FILELIST SNAPSHOT TableSnapshot(13, 2)
WHERE tag = 'data'; 
\end{lstlisting}

\noindent The client can also retrieve the list of manifest files to apply file pruning based on file-level statistics as follows.

\begin{lstlisting}
SELECT file_path 
FROM FILELIST SNAPSHOT TableSnapshot(13, 2)
WHERE tag = 'manifest'; 
\end{lstlisting}

\noindent Common data lakehouse operations such as CDC (Change Data Capture) extraction and in-memory snapshot update require identifying changes between versions. To this end, we provide \verb|DELTA| keyword for retrieving file-level additions and deletions between two snapshots. The query below demonstrates retrieving the delta between versions 2 and 3:

\begin{lstlisting}
SELECT action, file_path, tag
FROM FILELIST DELTA TableSnapshot(13, 2, 3); 
\end{lstlisting}

\noindent Raven is oblivious to the specific formats and types of files that a \verb|FILELIST| contains, allowing the system to be extended for new types of assets and storage formats in the future.

\section{System Architecture}
\label{sec:architecture}
\begin{figure}[t]
  \centering
  \includegraphics[width=\linewidth]{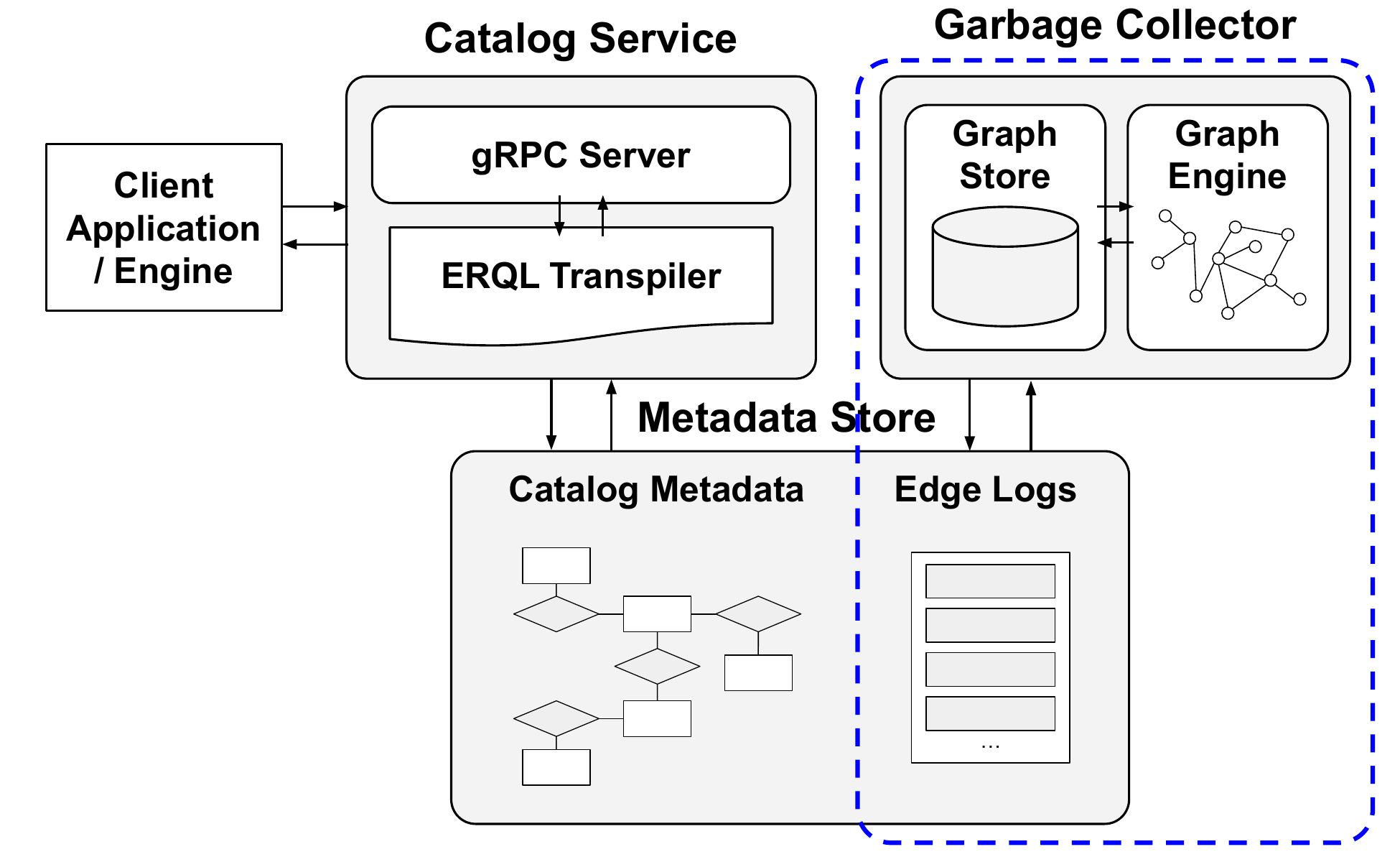}
  \caption{The high-level architecture of Raven.}
  \label{fig:overallarchitecture}
\end{figure}

In this section, we present Raven's system architecture and introduce two important features of Raven that address a subset of the OTF problems, including metadata bloat, high commit latencies, and the absence of multi-table transactions. Raven consists of three main components (Figure~\ref{fig:overallarchitecture}):

\begin{enumerate}[leftmargin=*,noitemsep,topsep=0pt]
    \item \textbf{Catalog Service:} This is the front-end server that handles requests from various clients via gRPC. 
    \item \textbf{Metadata Store:} The metadata store is the main backend data system that stores the metadata accessed by the catalog service and the garbage collector.
    \item \textbf{Garbage Collector} (Section~\ref{sec:gc})\textbf{:} The garbage collector periodically retrieves updates from the metadata store, identifies unreachable metadata objects and files, and deletes them.
\end{enumerate}

\noindent We now discuss the catalog service and the metadata store in detail, followed by the two key features: extended MoR (Merge-on-Read) and multi-table transactions.

\subsection{Catalog Service}
The catalog service provides a high-level gRPC~\cite{grpc} API, a high-performance RPC (Remote Procedure Call) framework developed by Google. Compute engines use this gRPC API to access and modify metadata, with multiple server instances deployable for scalability. Many functions, such as \textit{getTable()} and \textit{createTable()}, are CRUD operations implemented using ERQL, while \textit{execQuery()} allows clients to directly submit a limited set of ERQL queries for data-intensive operations (e.g., involving \verb|FILELIST|).

Inside the catalog service, a transpiler translates ERQL commands into the backend metadata store's native query language (SQL for our PostgreSQL prototype). Because ERQL is a high-level declarative language, there is no restriction on the choice of metadata store, and transpilers for other data systems may be added as plugin modules. In the ERQL-to-SQL translation scheme, each entity or relationship type maps to a table, and read queries translate into inner joins on object IDs. For abstract entity types, queries expand into a union of subqueries over derived concrete entity types. Because union scales optimization complexity linearly with subquery count (unlike a join), this avoids excessive query planning overhead.

Because the volume of file-level metadata remains substantial even after offloading statistics to OTFs, far exceeding other metadata types, we store it in a partitioned table grouped by \verb|FILELIST| history. Each partition stores file-level metadata with two additional fields, \textit{start\_vid} and \textit{end\_vid}, which determine the file's local lifetime within that \verb|FILELIST| history. Any query against \verb|FILELIST| translates into a query against the corresponding table partition with predicates on \textit{start\_vid} and \textit{end\_vid}. For each file, the metadata store also maintains a global \verb|FILELIST| count, tracking the number of \verb|FILELIST| histories in which the file is currently alive.

Lastly, second-order metadata, including entity and relationship type definitions and retention policies, is stored as catalog tables in the metadata store and cached in memory by each catalog service server.

\subsection{Metadata Store}
There is a wide range of OLTP (Online Transaction Processing) systems, including RDBMSs~\cite{postgres, mysql,mariadb,mssql}, graph databases~\cite{neo4j,neptune,janusgraph,tigergraph,zhu13livegraph}, column-family stores~\cite{bigtable,hbase,cassandra}, document stores~\cite{arangodb,mongodb}, and key-value stores~\cite{dynamodb,rocksdb,foundationdb}. The primary factor in selecting the underlying system was robust support for multi-row transactions with a strong isolation level, without which advanced features like multi-table transactions and garbage collection are difficult to implement correctly. Although we currently use PostgreSQL for its maturity and ease of development, other OLTP systems providing serializable transactions are viable alternatives.

While scalability is a valid concern, offloading file-level statistics to OTFs and adopting a constrained relational view alleviate pressure on the metadata store. In production, the compute engine or catalog can further reduce stress on the store by caching dataset snapshots and applying incremental updates via \verb|FILELIST DELTA| commands. Consequently, for a single enterprise, large-scale distributed systems like FoundationDB~\cite{foundationdb} or Spanner~\cite{spanner} would be required only in extreme circumstances. Fault tolerance is another important concern, which can be addressed using standard replication schemes or more specialized solutions, such as Amazon Aurora~\cite{verbitski2017amazon} and Neon~\cite{neon}.

\subsection{Base Transaction Protocol}

Before introducing the two features addressing OTF limitations, we describe Raven's core table-level concurrency control mechanism, which combines OCC (Optimistic Concurrency Control) with MVCC (Multi-Version Concurrency Control). A typical read-write path proceeds as follows:

\begin{enumerate}[leftmargin=*,noitemsep,topsep=0pt]
    \item \textbf{Snapshot Retrieval:} The client retrieves the latest table snapshot and its version ID from Raven. The table snapshot is retrieved from the metadata store in a single transaction.
    \item \textbf{Read Phase:} Via ERQL queries to Raven, the client retrieves the list of metadata and/or data files visible in the current table snapshot. The client reads the target files.
    \item \textbf{Write Phase:} The client writes new data and/or metadata files and sends the list of newly created or logically deleted files to Raven, which incrementally builds an ERQL command to apply these changes.
    \item \textbf{Raven Commit:} Upon invoking \textit{commit()}, the catalog service translates the ERQL command into SQL and executes it in the metadata store as a single serializable transaction, acting as a CAS operation on the target table snapshot. The new files become visible to other clients only after the transaction commits. If the target table snapshot has a higher version ID than what was initially retrieved, the transaction is aborted.
    \item \textbf{Post-Commit:} If \textit{commit()} aborts due to a transaction failure in the metadata store, the client handles recovery (e.g., retrying or deleting written files).
\end{enumerate}

\noindent Throughout the remainder of this paper, an OTF commit under Raven refers specifically to generating the relevant OTF files (delta log and checkpoint files for Delta Lake; manifest, manifest list, and metadata files for Iceberg). Thus, Iceberg omits its final catalog-level CAS operation, whereas Delta Lake still requires an atomic rename to create a new delta log file.

\subsection{Extended Merge-on-Read}

We next introduce extended MoR (Merge-on-Read) to reduce commit latency and metadata bloat. Extending the data-layer MoR strategy for small updates~\cite{apple_iceberg} to the metadata layer, extended MoR targets small, frequent updates where OTF commit latency and metadata file generation costs cannot be amortized as in bulk updates.

Because Raven already tracks file metadata in its transactional metadata store, it can defer OTF commits rather than eagerly generating OTF metadata files on every update. In the fast write path, the client follows the base OCC protocol to simply register new data files in Raven with a ``new data'' tag. Because these files are not yet tracked by the OTF, the read path merges them during scans. In the read path, after retrieving the table snapshot, the client issues ERQL queries to fetch both the OTF metadata files (the metadata file for Iceberg; the latest delta log file for Delta Lake) and all ``new data'' files visible in the current table snapshot. The client then reconstructs the OTF snapshot and merges its file list with the list of ``new data'' files.

To prevent the list of files subject to merge from growing indefinitely, periodic \textbf{metadata compaction} materializes ``new data'' files into OTF metadata, following the base OCC protocol. During the write phase, all ``new data'' files visible in the current snapshot are committed to the respective OTF. The resulting OTF files and tag updates (from ``new data'' to ``data'') for the target data files are then committed to Raven, making the changes visible. If the Raven commit aborts due to a conflict, the new OTF files are reclaimed safely as they are not visible until the final Raven commit succeeds. Performing OTF commits only periodically yields (1) fewer small metadata files and (2) lower average commit latency. Unlike batching, extended MoR preserves data freshness because Raven commits become visible prior to OTF materialization. However, because the metadata store lacks file-level statistics, ``new data'' files cannot be pruned before scanning even if they contain no relevant data, potentially impacting read performance. Currently, this feature supports only fast appends, but it is straightforward to extend it to updates and deletes.

\subsection{Multi-Table Transactions}

Raven also extends the base protocol to support multi-table OTF transactions under snapshot isolation. Like the base transaction protocol, all clients must go through Raven to avoid observing uncommitted changes.

\begin{enumerate}[leftmargin=*,noitemsep,topsep=0pt]
    \item \textbf{Snapshot Retrieval:} The client invokes \textit{startTransaction()} with target table names (both read and write tables) to retrieve their snapshots and version IDs from the metadata store in a single transaction.
    \item \textbf{Read Phase:} The client fetches the OTF metadata files (Iceberg metadata file; Delta Lake latest delta log) visible in the current table snapshots from Raven via ERQL queries, reconstructs the OTF snapshots, and scans the data files.
    \item \textbf{Write Phase:} The client writes new files and sends their paths to Raven, which constructs an ERQL command.
    \item \textbf{OTF Commit:} The client performs OTF commits, captures the newly generated OTF files, and sends them to Raven to append to the ERQL command.
    \item \textbf{Raven Commit:} The catalog service translates the ERQL command into SQL and submits it as a single serializable transaction to the metadata store, acting as a multi-row CAS across all modified table snapshots. New data and OTF files become visible only after this commit succeeds.
    \item \textbf{Post-Commit:} On abort, written data and OTF files are deleted. This cleanup is safe because uncommitted files are never visible to clients going through Raven.
\end{enumerate}

Committing metadata changes as a single serializable transaction guarantees a consistent cross-table snapshot. However, because Raven only applies the final catalog transaction to modified tables, it only detects ww-dependencies and permits write skew. If any OTF commit fails (e.g., during Delta Lake's atomic rename), the entire transaction aborts and deletes all newly created files. While the protocol does not currently handle orphan files from client crashes, a future remedy is a pre-commit step that allocates file names in a session-referenced temporary file list before file creation: moving them to active table file lists upon commit, or reclaiming them after the session ends. For Delta Lake, this mechanism is essential to prevent orphaned delta log files from blocking subsequent client commits.

\section{Garbage Collection}
\label{sec:gc}

Garbage collection is one of the most critical maintenance operations in a data lakehouse. Because data modifications continuously generate new files, older files and their associated metadata eventually become obsolete and must be reclaimed. Without strict safety guarantees, a garbage collector can delete active data, compromising data integrity. We first diagnose the limitations of garbage collection in existing OTFs.

\subsection{OTF Limitations}
\label{subsec:otfgclimit}
OTFs, including Delta Lake, Iceberg, and Hudi, primarily rely on retention-based garbage collection policies~\cite{delta_table_utility,iceberg_retention}, removing older table snapshots and their underlying data files once they exceed a configured time threshold. However, this compromises safety whenever older snapshots remain in active use. For instance, if an ETL pipeline accessing a snapshot runs longer than the retention window, the garbage collector may prematurely delete the snapshot's files, causing the job to fail. Furthermore, existing OTFs fail to account for access paths that span multiple tables or assets. For example, Delta Lake's \texttt{VACUUM} command reclaims physical files without checking for incoming references from shallow clones, risking corruption of dependent datasets~\cite{delta_table_utility}. Iceberg's \texttt{expireSnapshots()} procedure maintains local safety by computing the set difference between expired and active snapshots to identify unreferenced files. However, this safety guarantee relies on a restrictive constraint: cloning is confined to the scope of a single table directory path, precluding support for cross-table shallow clones.

\begin{figure}[t]
  \centering
  \includegraphics[width=\linewidth]{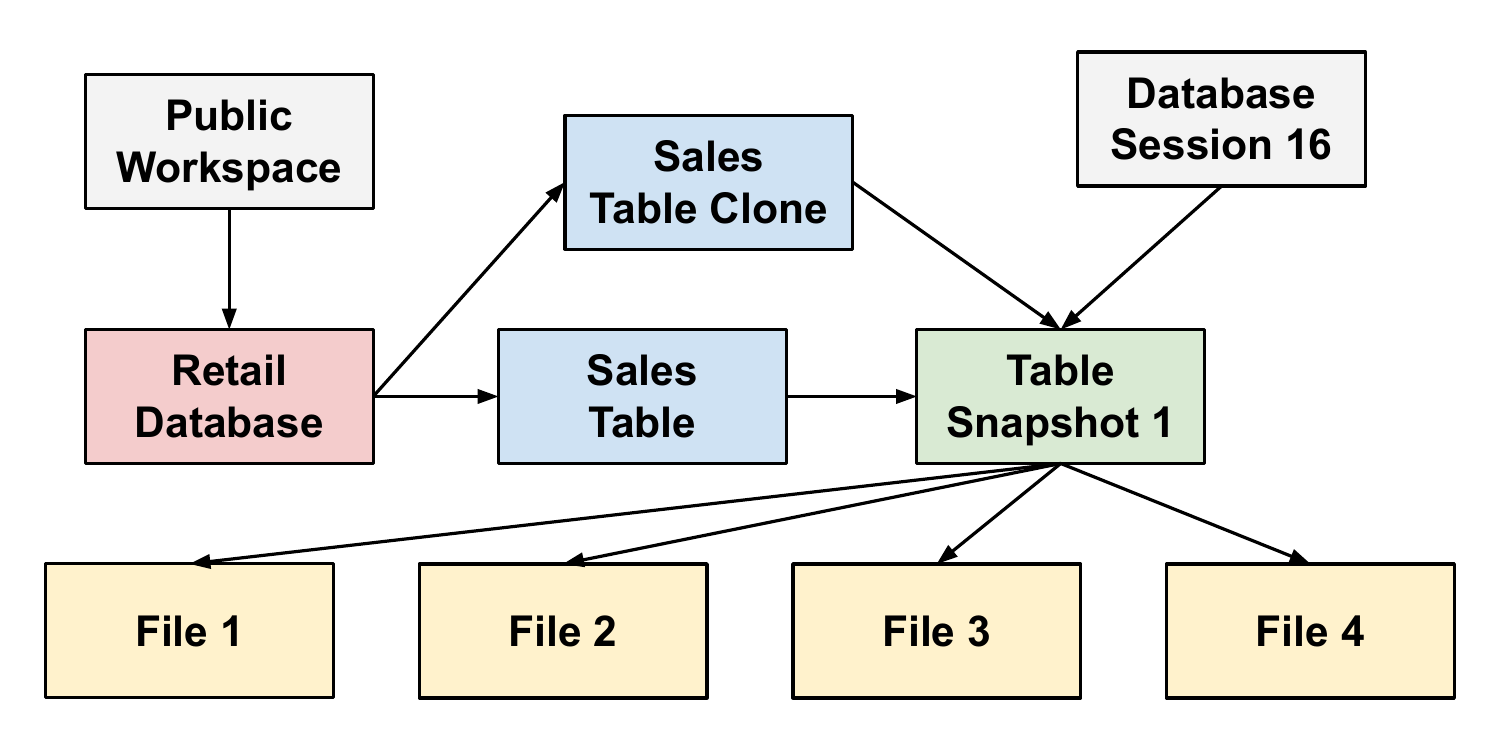}
  \caption{Graphical representation of example access paths, consisting of references between different entities.}
  \label{fig:validaccesspaths}
\end{figure}

Fundamentally, OTFs fail to track references across multiple file access paths. As illustrated in Figure~\ref{fig:validaccesspaths}, data files can be referenced through more than one distinct path:

\begin{itemize}[leftmargin=*,noitemsep,topsep=0pt]
    \item A workspace references a database, which references a table, which points to a snapshot that directly references data files.
    \item A cloned table references a snapshot in a separate base table, sharing the underlying data files.
    \item An active database session holds a reference to the table snapshot it is currently accessing.
\end{itemize}

\noindent Implementing precise reference tracking across these paths directly on top of OTFs is challenging because metadata is dispersed across multiple files in a cloud object store. Traversing these reference paths incurs substantial overhead due to high cloud object store latency. Furthermore, weak transactional guarantees make even straightforward mechanisms, such as reference counting, difficult to implement. Finally, the unstructured nature of native cloud storage APIs (key-value abstraction) further compounds these challenges.

\subsection{Problem Formalization}
To address the limitations of existing solutions from first principles, we formalize garbage collection over three object operations: $create$, $access$, and $delete$. Given an input schedule $s$ of $create$ and $access$ operations, the garbage collector must output a $delete$ schedule $g$ such that the combined schedule $s \oplus g$ is both (1) \textbf{complete}, where every object eventually has a $delete$ operation after its last $access$, and (2) \textbf{safe}, where each object's $delete$ succeeds its $create$ and all $access$ operations so no premature deletions occur.

Because garbage collection operates in real time, it requires a \textbf{liveness criterion} to determine whether an object will be accessed in the future. Existing OTFs rely on a \textbf{retention period} policy that assumes objects older than a threshold are obsolete, a heuristic that can fail and cause premature deletions. An alternative is \textbf{reachability}: modeling objects and their references as a directed graph, an object is considered alive only if a valid access path exists from a designated root set, allowing unreachable objects to be safely reclaimed. While OTFs fail precisely because they do not fully account for reachability, reachability alone is also insufficient in data lakehouses because accidental user operations (e.g., dropping a table) can prematurely delete references. Consequently, combining both policies is necessary to ensure safety.

\subsection{Referential Relationships}
\label{subsec:refrel}

To support garbage collection, we extend ERQL with \textbf{referential relationships} that natively incorporate reachability semantics into the metadata model (Figure~\ref{fig:validaccesspaths}). Because not every relationship is a reference, users must specify via DDL whether a relationship type is also a reference, as follows:

\begin{lstlisting}
ALTER RELATIONSHIP Contains(Database, Asset)
SET Database REFERENCES Asset;
\end{lstlisting}

\noindent For an access path to be considered valid, it must start from an entity belonging to a predefined \textbf{root set}. We call the corresponding entity types \textbf{root entity types}, which can be declared as shown below:

\begin{lstlisting}
ALTER ENTITY Workspace SET ROOT;
ALTER ENTITY Session SET ROOT;
\end{lstlisting}

To support retention-based recovery as mentioned above, we also allow users to configure a retention period for references. When a reference is dropped, it is marked rather than deleted for the specified duration. An entity reachable only via a path containing such a ``ghost'' reference is no longer visible, yet remains recoverable during its retention window and is deleted only after the retention period expires. For example, the default retention period for \verb|Contains(Database, Table)| can be set as follows:

\begin{lstlisting}
ALTER RELATIONSHIP Contains(Database, Table)
SET RETENTION INTERVAL '7 days';
\end{lstlisting}

\noindent The default retention period can be overridden with the \verb|OVERRIDING| keyword at the time a reference is dropped.

Raven defines an entity as alive if it is reachable from a root entity via a path of references, each of which is either (1) active or (2) deleted but unexpired. This constraint cannot be enforced by Raven alone; for example, once a compute engine retrieves the list of data files for a table snapshot, Raven cannot stop the engine from reading them after the last reference is dropped (e.g., by a \verb|VACUUM| operation). Instead, Raven provides an interface for clients to cooperate in enforcing this constraint. To safely access an asset snapshot, the client creates a root \textit{session entity} referencing the target snapshot so that it is not deleted even if other references are dropped. Abandoned session entities where a client fails to delete its session can be mitigated using leases with a Time to Live (TTL), reinforced by a heartbeat mechanism; cooperative clients locally track their lease and abort execution if it expires. Because clients in this architecture are internal execution engines and applications within the same enterprise lakehouse, we do not consider non-cooperative or adversarial behavior to be a realistic assumption.

Because file-level metadata is represented using \verb|FILELIST|, file-level references are implicit within \texttt{FILELIST} metadata. A file can span multiple \verb|FILELIST| histories, so its references are the union of all references across those \verb|FILELIST| histories. Accordingly, the metadata store maintains a global \verb|FILELIST| count for each file, representing how many \verb|FILELIST| histories (not \verb|FILELIST| versions) in which the file is currently alive.

We model the referential structure as a directed graph $G=(V,E)$, where $V$ is the set of entities and $E$ contains active or deleted, but unexpired, referential relationships. Given a root set $R$, the garbage collector identifies any vertex $v \in V$ unreachable from all $r \in R$. Because Raven uses reference counting, which cannot handle cycles, it prohibits DDL declarations that introduce referential cycles into the ER schema. This imposes minimal constraint in practice, as lakehouse referential structures are naturally acyclic, and non-referential relationship schemas remain unrestricted.

\subsection{Implementation}

\subsubsection{Overview}
Because the metadata store maintains the complete referential graph topology, including file-level references, transactionally under serializable isolation, it serves as a reliable source of truth for consistent garbage collection snapshots. However, executing graph traversals directly against the metadata store is problematic for two key reasons. First, graph algorithms inherently exhibit poor data locality due to irregular structural topologies and random access patterns~\cite{lumsdaine2007challenges,fuchs2022sortledton}. Second, garbage collection traversals require only graph connectivity and minimal algorithm-specific state (e.g., a mark bit per vertex). Consequently, direct traversal against the metadata store causes severe read amplification and random disk I/O. Furthermore, running these heavy traversal queries directly on the metadata store risks resource contention, potentially disrupting concurrent user operations.

Based on these considerations, we design the garbage collector as a specialized graph analytics engine that maintains a \textbf{dynamic graphical view} of the catalog metadata. The garbage collector periodically ingests updates from the metadata store, analyzes the graph to identify unreachable objects, and deletes them. The garbage collector encompasses the following three components (Figure~\ref{fig:overallarchitecture}):
\begin{enumerate}[leftmargin=*,noitemsep,topsep=0pt]
    \item \textbf{Metadata Store}: The metadata store is the single source of truth that stores the underlying metadata. It maintains a log of updates retrieved by the garbage collector.
    \item \textbf{Graph Store}: The graph store maintains an on-disk checkpoint of the graph state used for recovery after any failure. 
    \item \textbf{Graph Engine}: The graph engine maintains an in-memory graph state, on which the garbage collector operates to identify unreachable objects. The in-memory graph can handle random access patterns without performance issues.
\end{enumerate}

\noindent The graph store and graph engine are co-located on a dedicated machine, while the metadata store operates separately. 

\begin{figure}[t]
  \centering
  \includegraphics[width=\linewidth]{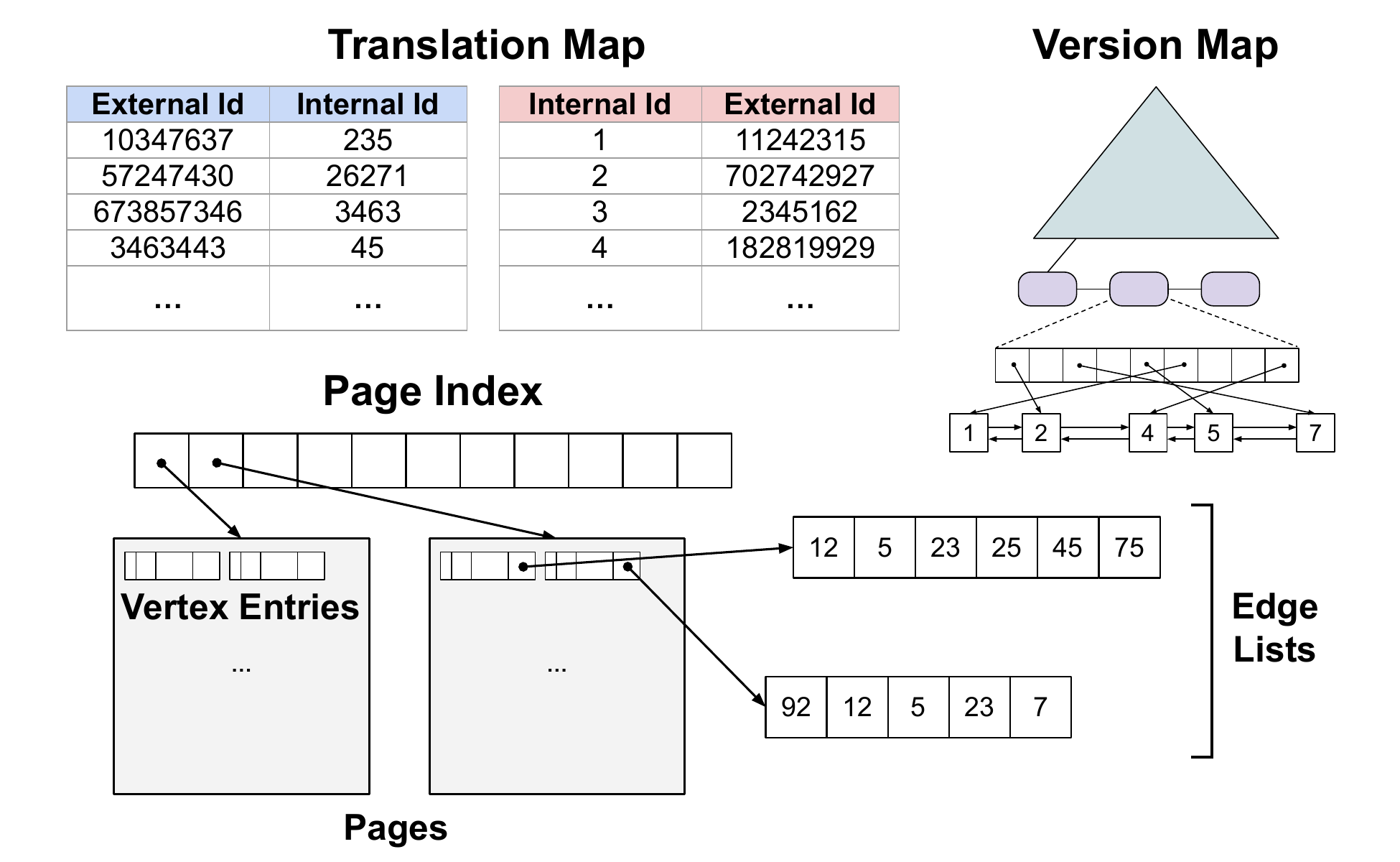}
  \caption{Main data structures of the graph engine.}
  \label{fig:gcarchitecture}
\end{figure}

\subsubsection{Main Process}
The metadata store logs changes to the object graph, additions and deletions of referential relationships, across two tables. Each log record is identified by source and destination object IDs (and version IDs for \verb|FILELIST|). Time is divided into epochs, with one garbage collection cycle per epoch. The main process initiates a cycle by incrementing the epoch ID, retrieving log records up to the previous epoch (all additions plus deletions whose retention expired), and updating the graph store. The garbage collector only requires knowing if an edge count is non-zero at an epoch boundary, so the graph store tracks only total addition and deletion counts per edge ID (and the epoch ID for durability). There may be multiple updates to the same edge, so the counts are aggregated in memory before being atomically applied to the graph store. After changes are committed, the corresponding log records in the metadata store are removed in a single transaction.

The in-memory graph engine receives the same updates as the graph store: additions first, followed by deletions. Since the system only requires correct graph state at epoch boundaries, updates can be reordered as long as an addition precedes its corresponding deletion, guaranteed by the update scheme. Figure~\ref{fig:gcarchitecture} illustrates the graph engine's main data structures. The \textbf{translation map} maintains a mapping between external object IDs and internal vertex IDs. The \textbf{page index} points to dynamically allocated pages of vertex records, where each record tracks a reference count, edge list pointer, and metadata. Edge lists use dynamic arrays when small and hash tables otherwise. The engine also maintains a free list of pages whose load factor is lower than a threshold. Lastly, the \textbf{version map} maps a \verb|FILELIST| object ID to its version chain, which keeps track of the list of live versions/snapshots.

Unlike tracing mechanisms, which require full graph traversals from the root set, the graph engine uses reference counting to perform a selective traversal starting from unreachable objects identified during updates. Starting with this initial active set, the garbage collector performs a traversal and in-place updates (Algorithm~\ref{alg:gc}) to locate unreachable objects and deallocate their in-memory vertex records. It then deletes the corresponding entities and relationships from the metadata store. Once these changes to the metadata store are committed, unreachable edge entries are deleted from the graph store to limit its size. Unlike references dropped by clients that are logged in the metadata store, references that become unreachable as a byproduct are deleted directly by the garbage collector, reducing data transfer overhead. The reachability of a root entity itself is tracked by an incoming reference from a virtual super root entity.

\begin{algorithm}[t]
\caption{Garbage Collection Algorithm}
\label{alg:gc}
\begin{algorithmic}[1]
    \STATE $D \gets \emptyset$
    \STATE \textbf{while} $A$ is not empty
        \STATE \hspace{2em} $v \gets $ an element from $A$
        \STATE \hspace{2em} $A \gets A \setminus \{ v \}$
        \STATE \hspace{2em} $D \gets D \cup \{ v \}$
        \STATE \hspace{2em} \textbf{for} $(v,w) \in v.outgoing\_edges$
            \STATE \hspace{4em} $w.ref\_count \gets w.ref\_count - 1$
            \STATE \hspace{4em} \textbf{if} $w.ref\_count = 0$
                \STATE \hspace{6em} $A \gets A \cup \{ w \}$
        \STATE \hspace{2em} // update reference graph
        \STATE \hspace{2em} $V \gets V \setminus \{ v \}$
        \STATE \hspace{2em} $E \gets E \setminus v.outgoing\_edges$
    \STATE \textbf{return} $D$
\end{algorithmic}
\end{algorithm}

\begin{figure*}[t]
  \centering
  \includegraphics[width=\linewidth]{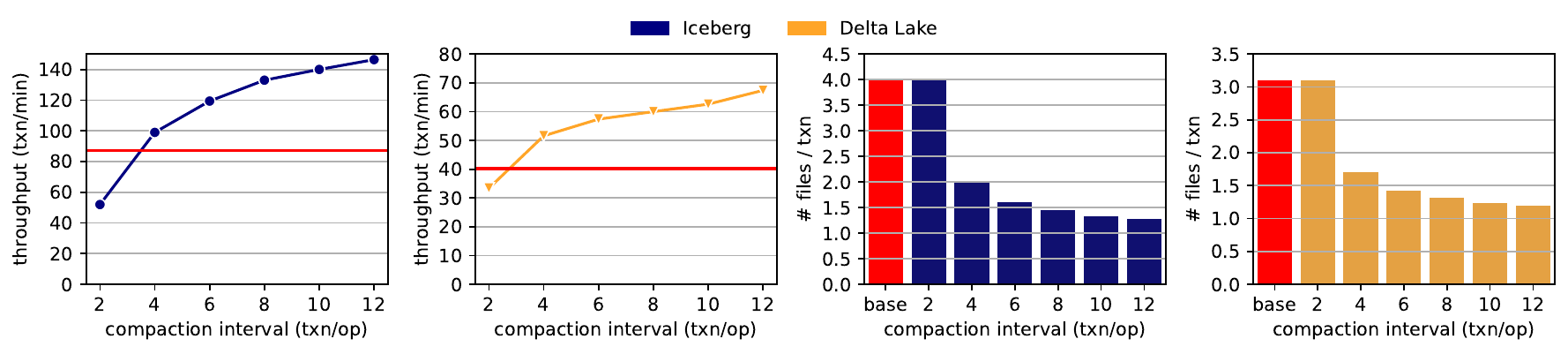}
  \caption{Throughput and total number of files created per transaction for the extended MoR implementation of Iceberg and Delta Lake across different compaction intervals. The baseline results without Raven integration are colored in red.}
  \label{fig:plot1}
\end{figure*}

A file is visible in a \verb|FILELIST| version if its version ID falls between \textit{start\_vid} and \textit{end\_vid}. To track these implicit file references without storing explicit relationships, the graph engine maintains a \textbf{version chain}, a hash table index and an ordered linked list of live version IDs, for each \verb|FILELIST| history. When a \verb|FILELIST| version is deleted, its entry in the chain is removed, and the global \verb|FILELIST| count of the files whose local lifetimes fall between neighboring versions is decremented. Background threads scan the metadata store for files with a global \verb|FILELIST| count of zero and delete them. By tracking files at \verb|FILELIST| granularity, rather than individually, the graph engine avoids managing billions of file-level metadata objects, enabling the entire graph structure (expecting tens of millions of other catalog objects at most) to fit in the memory of a single machine.

\subsubsection{Durability}
To ensure fault tolerance, the graph store maintains a checkpoint of the graph state in RocksDB~\cite{rocksdb}, a high-performance embedded key-value store, across two collections: edge additions and edge deletions. Because edges can be updated repeatedly, each collection maps edge IDs to the epoch ID of the latest operation and their total operation count. Because metadata store logs are deleted only after the graph store update commits, the graph state can be reconstructed from the metadata store even if a crash occurs during updates. For idempotency, log aggregates (grouped by epoch ID) are applied to the respective graph store collection only if their epoch ID exceeds the edge's latest recorded epoch ID during recovery. The graph engine is then rebuilt from the graph store before resuming garbage collection. Although the garbage collector may reattempt deletions performed before the failure, deletion operations on metadata and files are idempotent. To prevent double-decrementing the global \verb|FILELIST| count, the decrement is computed strictly from the files actually removed from the relevant file lists in a single transaction.

\section{Experimental Evaluation}
\label{sec:eval}

We conduct our experiments across four physical machines on CloudLab (Wisconsin)~\cite{cloudlab}, each equipped with a 10-core Intel Xeon Silver 4114 and 192~GiB of DDR4 memory interconnected via 25~Gbps Ethernet, using AWS S3~\cite{s3} in US East (Ohio) for primary data storage. The four machines respectively host: (1) the client application (Raven's gRPC client paired with either Iceberg 1.10 and DuckDB 1.4.4, or Delta Lake Kernel 4.0 and its default engine); (2) the Raven catalog service; (3) the Raven metadata store; and (4) the Raven garbage collector. For comparative evaluations against DuckLake, we deploy (1) DuckDB 1.4.4 paired with DuckLake 0.3 client; and (2) DuckLake backend (PostgreSQL).

\begin{figure}[t]
  \centering
  \includegraphics[width=\linewidth]{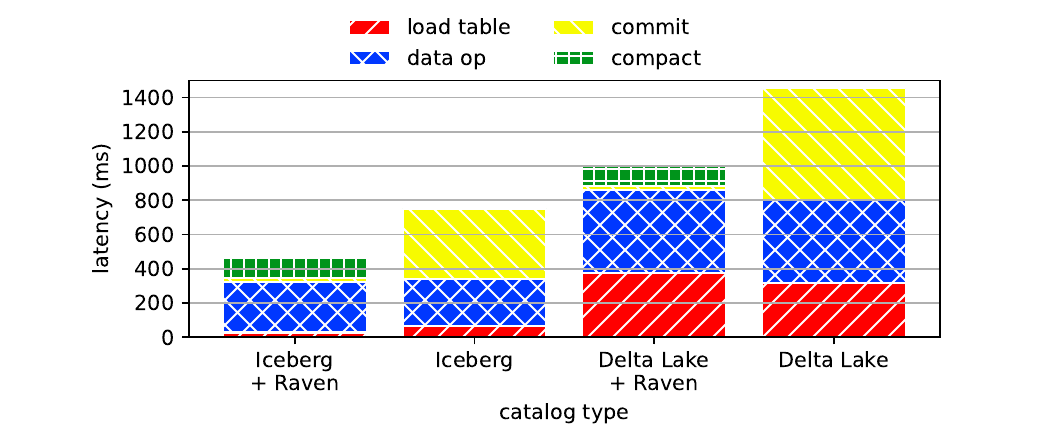}
  \caption{Mean latency breakdown of a single insert with and without extended MoR (compaction interval of 10 txn/op). Raven integration has lower combined commit and compact time than the commit time of OTF alone.} 
  \label{fig:plot2}
\end{figure}

\subsection{Extended Merge-on-Read}
\label{subsec:mor}

We evaluate how extended MoR reduces metadata bloat and commit latencies for Iceberg and Delta Lake. We use a microbenchmark where a single thread inserts data of a fixed size into the \textit{Store Sales} table of the TPC-DS benchmark.

\subsubsection{Compaction Interval} We first measure overall throughput and the number of files created while varying the compaction interval (the number of transactions per OTF commit) for both Iceberg and Delta Lake integrations (Figure~\ref{fig:plot1}). Compaction operation itself is a single transaction, so the compaction interval of $k$ results in a compaction per $k - 1$ inserts. Hence, with a compaction interval of 2, throughput is lower than the non-Raven baseline because metadata files are created and committed to Raven for every insert. However, as the interval increases, throughput improves significantly by amortizing metadata creation overhead. For example, at an interval of 10, Iceberg throughput reaches 140~txn/min, 61\% higher than the baseline (87~txn/min), while Delta Lake reaches 62~txn/min, 55\% higher than the baseline (40~txn/min). We also observe a substantial reduction in files created per transaction as the interval increases (lower metadata bloat). At an interval of 10, the number of files created per operation drops to 1.33 for Iceberg and 1.23 for Delta Lake, well below their respective baselines (4 and 3.1). As shown in Figure~\ref{fig:plot2}, the mean latency breakdown at an interval of 10 reveals that a lower amortized commit time is the primary contributing factor, as Raven's combined commit and compact latency is much lower than the commit latency of the baseline Iceberg and Delta Lake.

\begin{figure}[t]
  \centering
  \includegraphics[width=\linewidth]{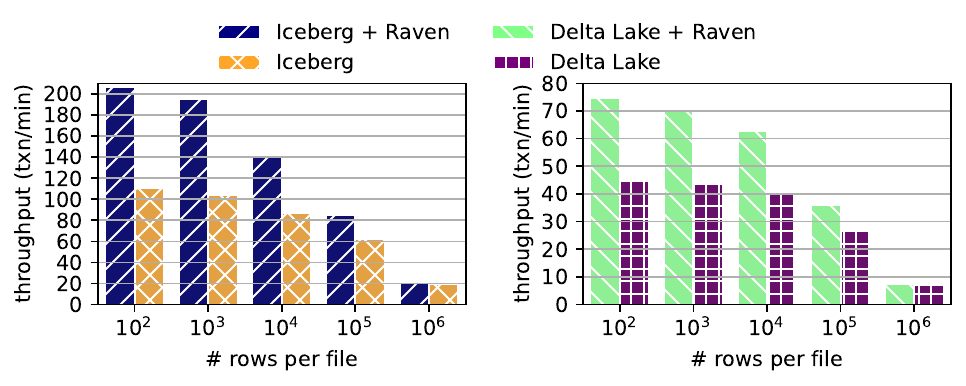}
  \caption{Insert transaction throughput for varying row counts per data file. Approximate file sizes are 14~KB, 96~KB, 0.9~MB, 9.2~MB, and 92~MB, respectively.} 
  \label{fig:plot3}
\end{figure}

\subsubsection{Data File Size} We next evaluate throughput as we increase the number of rows per data file by orders of magnitude. The results are shown in Figure~\ref{fig:plot3}. As expected, the throughput gap between the baseline OTFs and the Raven integrations narrows as data operation costs become dominant. However, even for data files with 100,000 rows (approximately 9.2~MB), Raven improves throughput by 36.8\% and 31.4\% for Iceberg and Delta Lake, respectively.

\subsubsection{Read Performance} 
\begin{figure}[t]
  \centering
  \includegraphics[width=\linewidth]{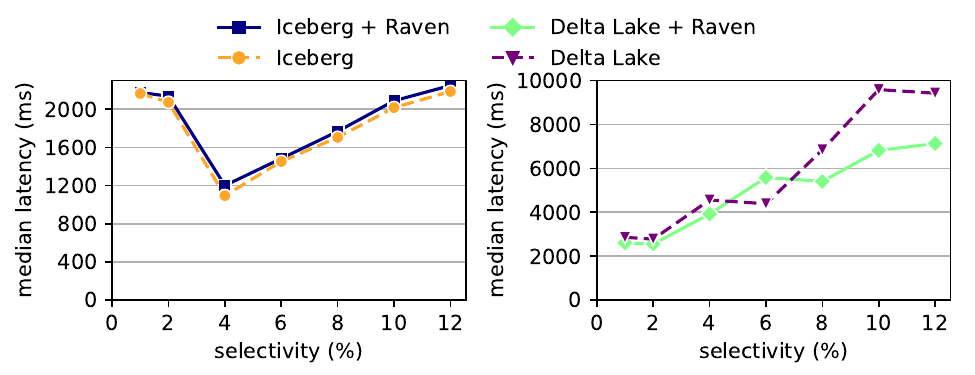}
  \caption{Median scan latency across varying selectivities with and without extended MoR.} 
  \label{fig:plot4}
\end{figure}

Finally, we show that extended MoR has minimal impact on Iceberg read performance. We populate tables with and without extended MoR (1,008 data files, 100,000 rows per file $\approx$9.2~MB, and a compaction interval of 10) and evaluate read queries across fixed selectivities (Figure~\ref{fig:plot4}). We deliberately chose the number of data files so that 7 files remain untracked by the OTF. For Iceberg, scan parallelism chosen by DuckDB during planning strongly influences runtime, as seen in the execution time drop from 2\% to 4\% selectivity. Latency differences between the two configurations remain small across selectivities, showing that the engine's parallel scan easily absorbs the overhead of scanning a few extra data files. For Delta Lake, read performance is governed by the Delta Kernel implementation rather than data file counts. Specifically, the Delta Kernel lazily fetches and parses log files from S3 during snapshot reconstruction, adding substantial latency. Variations in post-checkpoint log file counts mask the effect of scanning extra data files, making Delta Lake results inconclusive.

\subsection{Multi-Table Transactions}
\label{subsec:multi}

\subsubsection{OTF Integration} 
\begin{figure}[t]
  \centering
  \includegraphics[width=\linewidth]{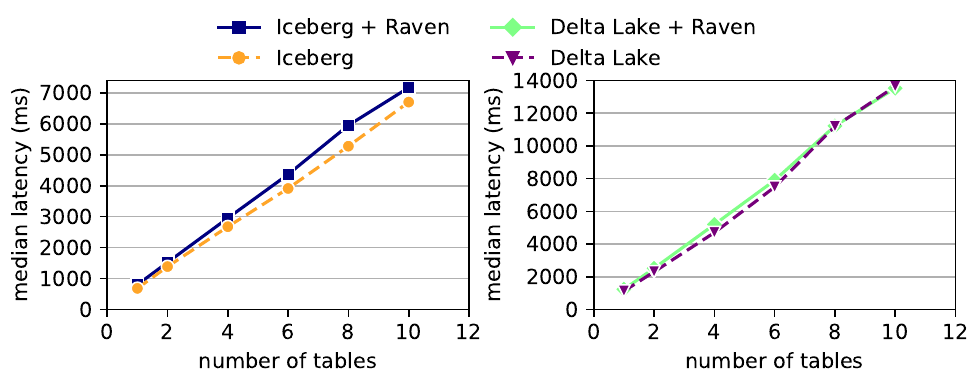}
  \caption{Median latency of a multi-table insert with and without a multi-table transaction across varying table counts.} 
  \label{fig:plot5}
\end{figure}

\begin{figure}[t]
  \centering
  \includegraphics[width=\linewidth]{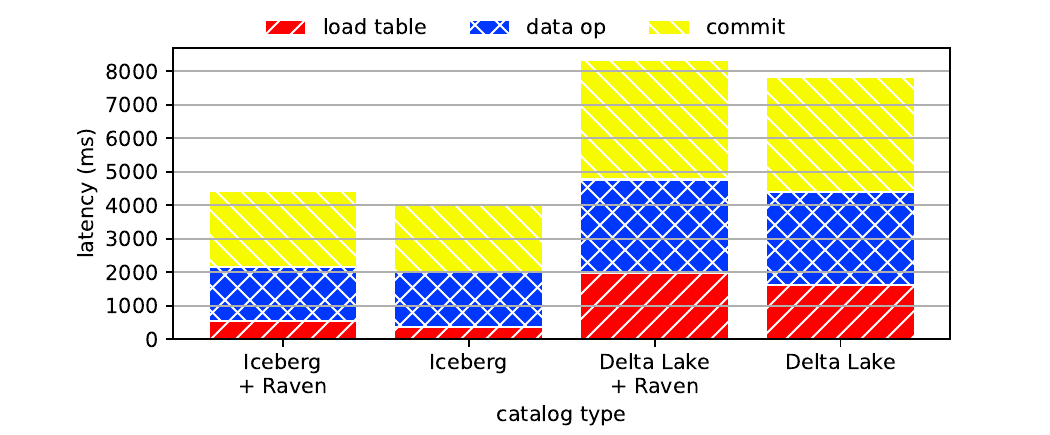}
  \caption{Mean latency breakdown of a multi-table insert into 6 tables with and without a multi-table transaction.} 
  \label{fig:plot6}
\end{figure}

Next, we evaluate the overhead incurred by Raven's multi-table transactions compared to sequential single-table transactions in baseline OTFs. Using 24 tables generated from the TPC-DS schema, our microbenchmark randomly selects a subset of tables and inserts 10,000 rows into each (Figure~\ref{fig:plot5}). For both Delta Lake and Iceberg, multi-table transactions introduce overhead from the per-table catalog RPCs required to fetch OTF metadata file paths. The mean latency breakdown for six-table transactions (Figure~\ref{fig:plot6}) reveals that Iceberg's overhead stems from both table loading and commits (176~ms and 279~ms across target tables, respectively), as each table is also refreshed during commit. On the other hand, Delta Lake's overhead stems primarily from table loading (366~ms) rather than commits (133~ms), which do not require a table refresh. Batching these catalog RPCs is a potential optimization for future work.

\subsubsection{DuckLake Comparison} 
\begin{figure}[t]
  \centering
  \includegraphics[width=\linewidth]{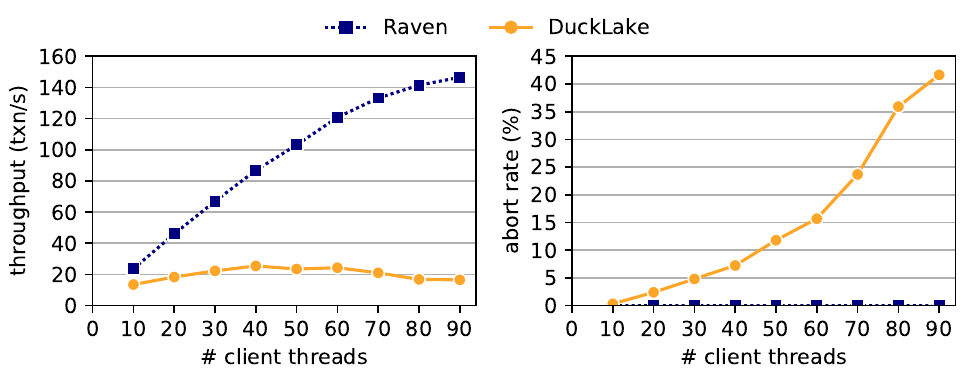}
  \caption{Throughput and abort rate of multi-table inserts across varying numbers of client threads. DuckLake's snapshot-based OCC causes higher contention.} 
  \label{fig:plot7}
\end{figure}

\begin{figure}[t]
  \centering
  \includegraphics[width=\linewidth]{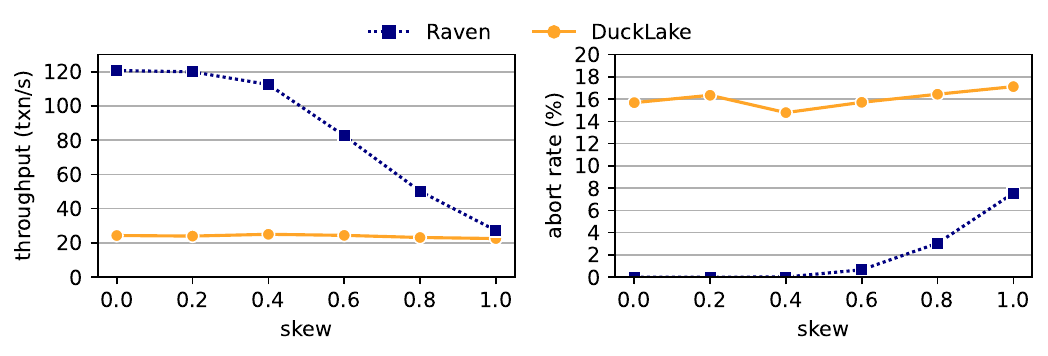}
  \caption{Throughput and abort rate of multi-table inserts across varying Zipfian skew factors (60 threads). Raven's performance converges to DuckLake's due to higher contention.} 
  \label{fig:plot8}
\end{figure}

We next compare Raven's catalog transaction path for multi-table transactions against DuckLake. To isolate the throughput and contention behavior of the catalog concurrency control mechanism, we minimize data overhead by omitting OTF integration and using lightweight single-column tables. Specifically, each transaction inserts a single row into 1 to 5 tables selected uniformly at random from a pool of 1,000 tables. As shown in Figure~\ref{fig:plot7}, Raven's throughput scales from 23.6~txn/s at 10 threads to 146.4~txn/s at 90 threads. In contrast, DuckLake reaches a maximum of 25.4~txn/s at 40 threads. Raven enforces table-level OCC, causing conflicts only when the target tables overlap. On the other hand, DuckLake serializes global catalog snapshot creation, creating a central bottleneck regardless of actual data conflicts. Retry validation logic prevents redundant execution when failures stem solely from catalog snapshot creation. But, failed snapshot creation still causes multiple retries and eventually leads to a transaction abort after 10 retries. As a result, DuckLake's abort rate reaches 41.6\% at 90 threads. For Zipfian distribution, DuckLake's throughput remains flat, whereas Raven's throughput drops toward DuckLake's baseline with increasing skew factor due to higher table-level contention (Figure~\ref{fig:plot8}).

\subsection{Statistics Overhead}
We show that Raven (with Iceberg) suffers substantially less statistics overhead than DuckLake using a single table of 100,000 single-row files across varying column counts.

\subsubsection{Metadata Footprint}
We measure PostgreSQL size to evaluate the catalog metadata footprint. As shown in Figure~\ref{fig:plot9}, DuckLake’s footprint scales linearly with column count, reaching 3.27~GB at 320 columns due to file-level statistics. In contrast, Raven's catalog storage size remains stable below 140~MB as it offloads file-level statistics to Iceberg.

\subsubsection{Read Performance}
\begin{figure}[t]
  \centering
  \includegraphics[width=\linewidth]{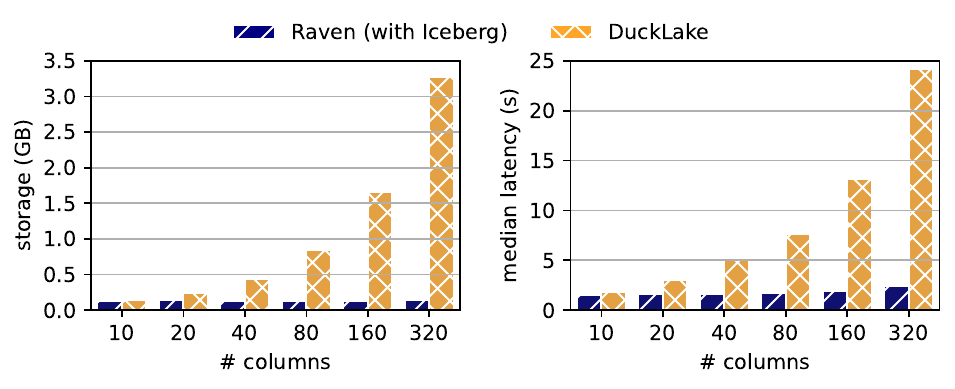}
  \caption{Catalog metadata footprint and single-row query latency across varying column counts. DuckLake suffers from increasing overhead due to file-level statistics.} 
  \label{fig:plot9}
\end{figure}

We measure query overhead by randomly selecting a single row from the table. Raven benefits from Iceberg, which can prune unnecessary manifest files and parallelize metadata loading, limiting its median latency increase from 1.82~s to 2.38~s in contrast to DuckLake's increase from 1.55~s to 24.17~s as column count increases from 10 to 320 (Figure~\ref{fig:plot9}). DuckLake retrieves file-level statistics from PostgreSQL using a full table scan with predicate filters. Because the physical volume of statistics scanned is proportional to the column count, DuckLake faces scalability issues.

\begin{figure}[t]
  \centering
  \includegraphics[width=\linewidth]{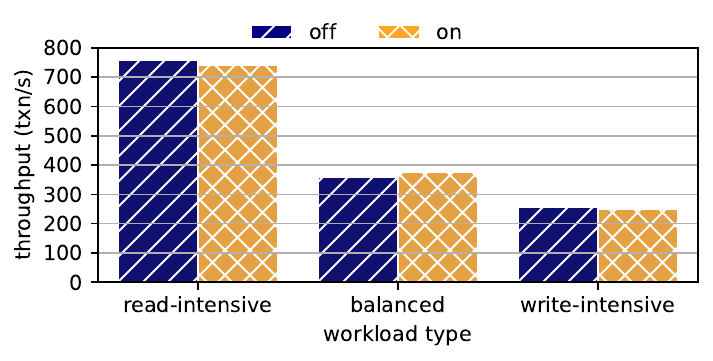}
  \caption{Throughput of different workloads with and without garbage collection, showing its minimal impact.} 
  \label{fig:plot10}
\end{figure}

\subsection{Garbage Collection}
\label{subsec:gc}
Lastly, we conduct an evaluation of the garbage collector. 

\subsubsection{Overhead} 
\begin{figure}[t]
  \centering
  \includegraphics[width=\linewidth]{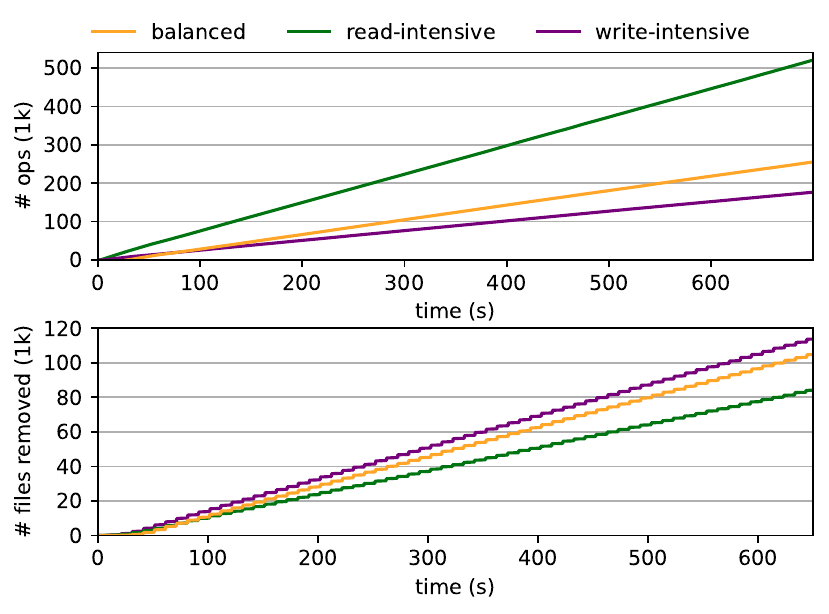}
  \caption{Total number of client operations and number of files removed as a function of time.} 
  \label{fig:plot11}
\end{figure}

We first demonstrate that garbage collection imposes minimal overhead on concurrent metadata operations. Across a database of 1,000 tables, 60 client threads execute three different types of operations: the \textit{refresh} operation is a read-only operation that scans new data files that have been added since the last operation; the \textit{insert} operation inserts new data into given tables; and the \textit{compact} operation (different from compaction in extended MoR) merges multiple new data files into a single data file and invokes \textit{expireSnapshots()}, retaining only the 10 most recent snapshots and triggering the garbage collector. To isolate GC overhead from transactional contention, each thread operates exclusively on its own assigned subset of tables. We evaluate three workloads with GC enabled and disabled: read-intensive (80\% read, 20\% write), balanced (50\% read, 50\% write), and write-intensive (20\% read, 80\% write), using a 10-second epoch period. Figure~\ref{fig:plot10} illustrates the main result: overall performance shows negligible difference regardless of GC activity. Figure~\ref{fig:plot11} shows operation throughput and file deletions over time. Metadata operation rates remain steady, while file deletions follow a ``jagged'' pattern due to batch removal at the end of each epoch.

\begin{figure}[t]
  \centering
  \includegraphics[width=\linewidth]{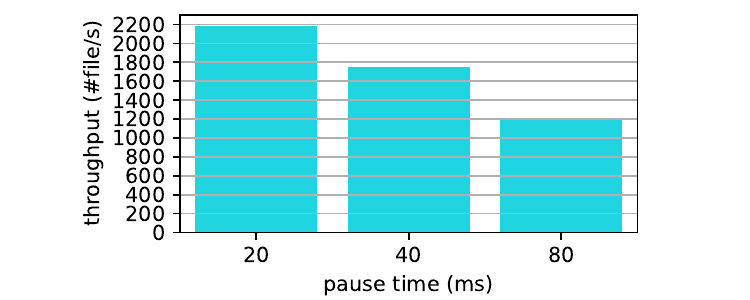}
  \caption{Rate of file removal for different client pause times.} 
  \label{fig:plot12}
\end{figure}

\subsubsection{Scalability} 
\begin{figure}[t]
  \centering
  \includegraphics[width=\linewidth]{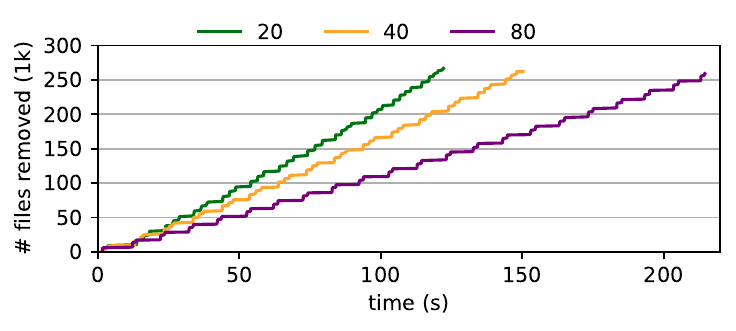}
  \caption{Total number of files removed by the garbage collector over time for different client pause times.} 
  \label{fig:plot13}
\end{figure}

Finally, we evaluate the scalability of the garbage collector. We first populate 1,000 tables with 300,000 data files using a mix of insert and compact operations. Next, 30 client threads select tables from their assigned groups and expire 10 snapshots per operation until there are no snapshots left to expire. In this setup, the GC workload generated per operation far exceeds client processing time. Consequently, executing snapshot expirations in a tight loop would exceed S3's rate limit (3,500 DELETEs/s per partitioned prefix), triggering throttling. We therefore regulate client throughput by introducing a pause interval between consecutive operations. Figure~\ref{fig:plot12} reports file deletion rates across three pause times (20~ms, 40~ms, and 80~ms). At a 20~ms pause, the lowest interval before hitting S3 limits, the system deletes 2,183.6~files/s. To verify that GC throughput is mainly constrained by S3 API limits and overheads, we repeated the experiment on a local filesystem with zero pause time, reaching 3,256.8~files/s. Lastly, Figure~\ref{fig:plot13} shows cumulative file deletions over time, following a near-linear trajectory with a ``jagged'' pattern caused by epoch-based bulk removals.

\section{Related Work}
\label{sec:relatedworks}

Prior work has also addressed multi-table transactions in data lakehouse systems. Nessie~\cite{nessie} uses Git-like branching semantics to isolate multi-table modifications before atomically merging them into the main branch. DuckLake~\cite{ducklake} employs a client-side OCC mechanism that optimistically generates a global catalog snapshot, executing conflict validation if the RDBMS transaction fails. LakeVilla~\cite{lakevilla} provides a pluggable module for Iceberg and Delta Lake, in which the client places marker files to lock table snapshots and track cross-table dependencies and commits to a global version log for cross-table consistency. TreeCat~\cite{treecat} supports multi-table transactions based on fine-grained optimistic predicate locking. Finally, Databricks announced support for multi-table transactions, but it remains restricted to proprietary compute engines integrated with Unity Catalog as of September 2026~\cite{databricks_txn}.

The garbage collection problem persists across many different domains. In programming languages, automatic heap management~\cite{jones2023garbage} relies primarily on global tracing (e.g., mark-sweep) or reference counting, employing advanced techniques, including concurrent~\cite{yuasa1990real,dijkstra,steele,baker} and generational garbage collection, and deferred~\cite{deutsch} and cyclic reference counting~\cite{bacon2001concurrent}. The MVCC mechanism of database systems requires garbage collection to remove tuple versions no longer visible to any active transactions, performed in the background~\cite{postgresql_vacuum} or as part of query processing~\cite{bottcher19,larson2013hekaton}. Partitioning schemes that allow efficient bulk deletion have also been devised~\cite{neumann2015fast,kim2022diva}.

\section{Conclusion}
\label{sec:conclusion}

We addressed the core architectural dilemma of data lakehouse metadata management. While file-based OTFs scale analytical workloads, they suffer from metadata bloat, high commit latencies, absence of multi-table transactions, and unsafe garbage collection. Database-only catalogs bottleneck on file-level statistics. To bridge this divide, we presented Raven, a catalog whose hybrid design isolates transactional control metadata from analytical metadata. Raven enables extended MoR to curb metadata bloat, coordinates multi-table transactions, and models asset lifecycles via an extensible ER abstraction for safe garbage collection. Our evaluation confirms that Raven lowers commit latency and metadata growth for Apache Iceberg and Delta Lake, maintains catalog storage orders of magnitude smaller than DuckLake, and deletes obsolete files at scale with minimal disruption.

Looking ahead, several important challenges remain. Given the open architecture of data lakehouses, \textbf{access control} is harder to address than in closed systems such as data warehouses. Although enterprise solutions such as Unity Catalog exist~\cite{unityaccess}, enforcing fine-grained policies such as row-level filters across different engines remains difficult. Second, we have not evaluated Raven at \textbf{an extreme scale}, which would likely require a metadata store backed by a highly scalable transactional system. Finally, \textbf{optimizing the mapping} from the ER model to the underlying storage remains an open opportunity to further improve performance by appropriately selecting a table representation for the metadata.

\begin{acks}
This research was supported by Dolby Laboratories.
\end{acks}

\section*{AI-Generated Content Acknowledgment}
The authors used Google Gemini to improve the grammatical fluency and readability of the manuscript. The authors also used GitHub Copilot and Google Gemini for coding and script development assistance.

\balance
% Use standard plain/abbrv bibliography to avoid missing bst errors on arXiv
\bibliographystyle{abbrv}
\bibliography{ref}

\end{document}